\documentclass[aps,prd,showpacs,showkeys,amsmath,amssymb]{revtex4}
\usepackage{graphicx}
\usepackage{dcolumn}
\usepackage{bm}
\begin{document}

\title{Pentaquark Magnetic Moments In Different Models}
\author{Y.-R. Liu}
\author{P.-Z. Huang}
\author{W.-Z. Deng}
\author{X.-L. Chen}
\affiliation{%
Department of Physics, Peking University, BEIJING 100871, CHINA}
\author{Shi-Lin Zhu}
\email{zhusl@th.phy.pku.edu.cn}
\affiliation{%
Department of Physics, Peking University, BEIJING 100871, CHINA\\
The Key Laboratory of Heavy Ion Physics, Ministry of Education,
Beijing 100871, China}

\date{\today}


\begin{abstract}
We calculate the magnetic moments of the pentaquark states from
different models and compare our results with predictions of other
groups.
\end{abstract}



\maketitle

\pagenumbering{arabic}

\section{Introduction}\label{sec1}

LEPS Collaboration first discovered a very narrow baryon state
around $1540$ MeV with strangeness $S=+1$, which was confirmed by
DIANA \cite{diana}, CLAS \cite{clas,clasnew}, SAPHIR
\cite{saphir}, ITEP \cite{itep} Collaboration. The positive
strangeness means that the valence quark number inside this baryon
is five. Such a state is definitely beyond the conventional quark
model. Now it is called the $\Theta^+$ pentaquark. Later, NA49
Collaboration found evidence for the existence of another narrow
baryon resonance with $Q=-2, S = -2, I = 3/2$ and mass of
$(1.862\pm 0.002) $ GeV/c$^2$ from high energy collision
experiment \cite{na49}. Experimentally there is evidence that
$\Theta^+$ particle may be an isoscalar from the featureless $p
K^+$ spectrum \cite{clas,saphir, clasnew}. The parity and angular
moment of the $\Theta^+$ particle have not been determined
experimentally.

One should keep in mind that Quantum Chromodynamics (QCD) as the
underlying theory of the strong interaction does allow the
possible existence of the non-conventional hadrons with the quark
content other than $q\bar q$ or $q q q$. Some examples are
glueballs ($gg, ggg, \cdots$), hybrid mesons ($q\bar q g$), and
other multi-quark states ($qq\bar q \bar q$, $qqqq\bar q$,
$qqq\bar q \bar q \bar q$, $qqqqqq, \cdots$). In fact, hybrid
mesons are found to mix freely with conventional mesons in the
large $N_c$ limit \cite{tom}. However, despite extensive
experimental searches in the past two decades, none of these
states has been firmly established until the surprising discovery
of pentaquarks \cite{pdg}.

These experiments have triggered heated discussions of the
interpretation of these resonances and their quantum numbers
\cite{page}-\cite{buddy}. The partial motivation of the recent
experimental search of the pentaquark state came from the work by
Diakonov et al. \cite{diak}. They proposed the existence of the
$J^P={1\over 2}^+$ $\Theta^+$ pentaquark at $1530$ MeV with a very
narrow width in the chiral soliton model (CSM).

However, such a picture is challenged by Refs.
\cite{page,cohen,princeton}. It was pointed out in \cite{page}
that identifying $N(1710)$ as a member of the anti-decuplet in the
CSM is kind of arbitrary. Using a more reasonable value for the
$N(1710)$ decay width as the input may lead to a big width of
$\Theta^+$ particle exceeding the present experimental upper
bound. Another serious challenge to the chiral soliton model came
from the large $N_c$ consistency consideration by Cohen
\cite{cohen}, in which he concluded that the prediction for
$\Theta^+$ properties based on collective quantization of CSM was
not valid. The reason is that collective quantization is
legitimate only for excitations which vanish as $N_c \to \infty$.
In Ref. \cite{princeton} it was shown that the exotic state may be
an artifact of the rigid rotator approach to the Skyrme model for
large $N_c$ and small $m_K$. Moreover, the mass of the pentaquark
state with the quark content $(d s d s \bar u)$ is rigourously
predicted in the chiral soliton model to be $2070$ MeV, which is
$210$ MeV higher than the experimental value measured by NA49
Collaboration \cite{na49}.

Although it is very difficult to obtain a low-mass $\Theta^+$
pentaquark using the picture of constituent quarks, it is still
possible to accommodate such a low mass in other theoretical
approaches. For example, the $\Theta^+$ pentaquark mass from QCD
sum rule formalism \cite{zhu,qsr} and lattice QCD simulation
\cite{lattice} is consistent with the experimental value. Treating
the pentaquark as the bound state of two tight bound diquarks and
one anti-quark or as the bound state of a diquark and a triquark
may also lower its mass \cite{jaffe,lipkin,shuryak}.

Because of its low mass, most of theoretical work postulated its
angular momentum to be one half. But it was pointed out that the
possibility of $J={3\over 2}$ still can not be excluded completely
\cite{close}. The parity of the $\Theta^+$ pentaquark is still a
big puzzle. Up to now theoretical papers in the literature can be
roughly classified into two types according to their assumption of
$\Theta^+$ parity. It's interesting to note that QCD sum rule
approach \cite{zhu,qsr}, lattice QCD calculation \cite{lattice}
and some models \cite{zhang,carlson,wu} favor the negative parity.
Many other theoretical papers assume positive parity for
$\Theta^+$ pentaquark as in the original Ref. \cite{diak}.
Recently some typical examples are the clustered quark models
proposed by Jaffe and Wilczek \cite{jaffe}, Shuryak and Zahed
\cite{shuryak}, and by Karliner and Lipkin \cite{lipkin}.

The $\Theta^+$ parity is very challenging. But the more
challenging one to theorists is the extreme narrow width of the
$\Theta^+$ particle. This baryon has such a narrow width that all
experiments can only set an upper bound. There is even stringent
limit arising from recent analysis that its width is less than
several MeV \cite{nuss,arndt,krein,cahn}.

Experience with conventional excited hadrons shows that their
widths are around one hundred MeV or even bigger if they lie 100
MeV above threshold and decay through S-wave or P-wave. For
comparison, the $S=-1$ hyperon $\Lambda (1520)$ $D_{03}$ state is
in the same mass region as the $\Theta^+$ pentaquark. Its angular
momentum and parity is $J^P={3\over 2}^-$. Its dominant two-body
decay is of D-wave with final states $N\bar K$, $\Sigma \pi$. With
a smaller phase space and higher partial wave, the width of
$\Lambda (1520)$ is $15.6\pm 1.0$ MeV \cite{pdg}. In contrast,
$\Theta^+$ decays through either S-wave or P-wave with a total
width less than several MeV, corresponding to negative or positive
parity respectively. Such a comparison is really disturbing.

If these states are further established and confirmed to have such
a narrow width, there must be a mysterious selection rule which is
absent from the conventional hadron interaction. Uncovering and
understanding such an underlying selection rule would be a top
priority to theorists.

We have estimated the mass of the pentaquark state with QCD sum
rules and found that the lowest pentaquark state lies around
$(1.55\pm 0.15)$ GeV \cite{zhu}. The baryon magnetic moment is
another fundamental observable as its mass, which encodes
information of the underlying quark structure and dynamics. We
have also employed the light cone QCD sum rules to extract the
magnetic moment of the $\Theta^+$ particle and found its value is
quite small \cite{huang}.

A first-principle dynamical determination of the pentaquark
quantum numbers is still not feasible right now. In this paper, we
shall start from different models available in literature and
study their phenomenological prediction, namely the magnetic
moments of pentaquarks. We shall calculate the pentaquark magnetic
moments in several typical models. They include (1) Jaffe and
Wilczek's model \cite{jaffe}, (2) Shuryak and Zahed's model
\cite{shuryak}, (3) Karliner and Lipkin's model \cite{lipkin}, (4)
Strottman's model \cite{strot}. Different models generally yield
different values. This in turn will affect the photo-production
and electro-production cross sections of pentaquarks. Such a study
will deepen our knowledge of pentaquark states and may help us
explore its dynamics and distinguish so many models in the
literature.

Our paper is organized as follows: Section I is an introduction. A
brief review of this field is presented. In Section II-III we
calculate pentaquark magnetic moments in the Jaffe and Wilczek's
model and Shuryak and Zahed's model respectively. In Section IV we
perform the same calculation in Lipkin and Karliner's model
treating pentaquarks as a bound state of a diquark and triquark.
In Section V we perform the calculation in Strottman's model,
treating pentaquarks as a cluster of four quarks and one
anti-quark. Our discussions are given in Section VI, where we also
compare our result with other groups' prediction.

\section{Pentaquark As A Bound State of Two Scalar Diquarks and One Anti-quark}\label{sec2}

Jaffe and Wilczek (JW) suggested that the observed $\Theta^+$
state is composed of an anti-strange quark and two highly
correlated up and down quark pairs arising from strong color-spin
correlation \cite{jaffe}. They treated the diquarks as point-like
particles obeying Bose statistics and predicted the isospin $3/2$
multiplet of $\Xi_5^{--}$ $(ddss \bar u)$ with $S=-2$ and
$J^P={1\over 2}^{+}$ around 1750 MeV. Such a state with the same
quantum number was observed by NA49 but with a much higher mass at
$1860$ MeV \cite{na49}.

According to JW's model, there will be a $J=\frac{1}{2}$ $SU_f(3)$
antidecuplet and an octet. But some pentaquarks in the octet may
mix with those in the antideculplet. Three pentaquarks have "pure"
components: $\Theta^+(udud\bar{s})$, $\Xi_5^+(usus\bar{d})$ and
$\Xi_5^{--}(dsds\bar{u})$. $\Theta^+$ is at the apex of the
antidecuplet and the other two are at the bottom corners. They
also predicted their corresponding masses of these exotic baryons.
In this section, we will calculate the magnetic moments of these
pentaquarks in this model.

The diquark arises from the strong chromomagnetic attraction or
instanton interaction between up quark and down quark when they
are in the color $\bf{\bar{3}}$ channel. In JW's model, each
diquark is in the antisymmetric $\bf{\bar{3}_c}$ color
representation.  The spin wave function of a scalar diquark is
antisymmetric and the space wave function is symmetric. Thus the
scalar diquark has positive parity. Because of the Pauli
principle, the total wave function of the diquark must be
antisymmetric. So the diquark must be in the antisymmetric
$\bf{\bar{3}_f}$ (Table \ref{tab1}). We may treat a $[ud]$-diquark
as an "effective" $s$-antiquark, $[ds]$-diquark as $u$-antiquark,
and $[su]$-diquark as $d$-antiquark in flavor and color symmetry
$SU(3)_f \otimes SU(3)_c$.

\begin{table*}[h]
\begin{center}
\begin{tabular}{ccccc}\hline
($Y_1,I_1$)&($Y_2,I_2$) & ($Y,I$) & $I_3$&Flavor Wave Functions\\
\hline
($\frac{1}{3},\frac{1}{2}$)&($\frac{1}{3},\frac{1}{2}$)&($\frac{2}{3},0$)&0&$[ud]=\frac{1}{\sqrt{2}}(ud-du)$\\
($\frac{1}{3},\frac{1}{2}$)&(-$\frac{2}{3}$,0)&(-$\frac{1}{3},\frac{1}{2}$)&$\frac{1}{2}$&$[us]=\frac{1}{\sqrt{2}}(us-su)$\\
(-$\frac{2}{3}$,0)&($\frac{1}{3},\frac{1}{2}$)&(-$\frac{1}{3},\frac{1}{2}$)&-$\frac{1}{2}$&$[ds]=\frac{1}{\sqrt{2}}(ds-sd)$\\
\hline
\end{tabular}
\end{center}
\caption {Diquark flavor wave functions. $Y$, $I$ and $I_3$ are
hypercharge, isospin and the third component of isospin
respectively. The subscripts 1 and 2 represent two quarks inside
the diquark. }\label{tab1}
\end{table*}

The total wave function of the two diquark system must be
symmetric under the diquark exchange. The color part is
antisymmetric. The spin wave function is certainly symmetric in
the case of scalar diquarks. Since the $\Theta^+$ particle is
assumed to be an isoscalar, the diquark-diquark system in
$\Theta^+$ also has $I=0$. Its $SU_f(3)$ flavor wave function is
symmetric $\bf{\bar{6}_f}$. So the spatial wave function of
diquark-diquark is antisymmetric, which implies the orbital
angular momentum between the two diquarks is odd. Thus the parity
of the two diquark system is negative. Finally the two diquarks
combine with an antiquark to form the exotic pentaquark state. The
total angular momentum and parity of the pentaquark will be
$J^P={1\over 2}^+$ or ${3\over 2 }^+$. In general, lower angular
momentum implies lower mass. So $J^P={1\over 2}^+$ is adpoted in
Ref. \cite{jaffe}. Pentaquarks belong to either the flavor octet
$\bf{8_f}$ or an anti-decuplet $\bf{\bar{10}_f}$ in JW's model.
The resulting flavor wave functions are listed in Table
\ref{tab2}.

\begin{table*}[h]
\begin{center}
\begin{tabular}{cc||cc}\hline
($Y,I,I_3$)       &$\bf{\bar{10}}$ &($Y,I,I_3$)&$\bf{8}$\\
\hline
(2,0,0)            &$[ud]^2\bar{s}$ &$-$&$-$\\
(1,$\frac{1}{2}$,$\frac{1}{2}$)&$\sqrt{\frac{2}{3}}[ud][us]_+\bar{s}+\sqrt\frac{1}{3}[ud]^2\bar{d}$
     &(1,$\frac{1}{2}$,$\frac{1}{2}$)&$\sqrt\frac{1}{3}[ud][us]_+\bar{s}-\sqrt{\frac{2}{3}}[ud]^2\bar{d}$\\
(1,$\frac{1}{2}$,-$\frac{1}{2}$)&$\sqrt{\frac{2}{3}}[ud][ds]_+\bar{s}+\sqrt\frac{1}{3}[ud]^2\bar{u}$
    &(1,$\frac{1}{2}$,$\frac{1}{2}$)&$\sqrt\frac{1}{3}[ud][ds]_+\bar{s}-\sqrt{\frac{2}{3}}[ud]^2\bar{u}$\\
(0,1,1)
&$\sqrt{\frac{2}{3}}[ud][us]_+\bar{d}+\sqrt\frac{1}{3}[us]^2\bar{s}$
     &(0,1,1)&$\sqrt{\frac{1}{3}}[ud][us]_+\bar{d}-\sqrt\frac{2}{3}[us]^2\bar{s}$\\
(0,1,0) &$\sqrt{\frac{1}{3}}([ud][ds]_+\bar{d}+[ud][us]_+\bar{u}$
     &(0,1,0)&$\sqrt{\frac{1}{6}}([ud][ds]_+\bar{d}+[ud][us]_+\bar{u})$\\
  &$+[us][ds]_+\bar{s})$&&$-\sqrt{\frac{2}{3}}[us][ds]_+\bar{s}$\\
(0,1,-1)
&$\sqrt{\frac{2}{3}}[ud][ds]_+\bar{u}+\sqrt\frac{1}{3}[ds]^2\bar{s}$
     &(0,1,-1)&$\sqrt{\frac{1}{3}}[ud][ds]_+\bar{u}-\sqrt\frac{2}{3}[ds]^2\bar{s}$\\
(-1,$\frac{3}{2}$,$\frac{3}{2}$)&$[us]^2\bar{d}$ &$-$&$-$\\
(-1,$\frac{3}{2}$,$\frac{1}{2}$)&$\sqrt{\frac{2}{3}}[us][ds]_+\bar{d}+\sqrt\frac{1}{3}[us]^2\bar{u}$
     &(-1,$\frac{1}{2}$,$\frac{1}{2}$)&$\sqrt{\frac{1}{3}}[us][ds]_+\bar{d}-\sqrt\frac{2}{3}[us]^2\bar{u}$\\
(-1,$\frac{3}{2}$,-$\frac{1}{2}$&$\sqrt{\frac{2}{3}}[ds][us]_+\bar{u}+\sqrt\frac{1}{3}[ds]^2\bar{d}$
     &(-1,$\frac{1}{2}$,-$\frac{1}{2}$)&$\sqrt{\frac{1}{3}}[ds][us]_+\bar{u}-\sqrt\frac{2}{3}[ds]^2\bar{d}$\\
(-1,$\frac{3}{2}$,-$\frac{3}{2}$)&$[ds]^2\bar{u}$ &$-$&$-$\\
$-$&$-$&(0,0,0)&$\sqrt\frac{1}{2}([ud][ds]_+\bar{d}-[ud][us]_+\bar{u})$\\
 \hline
\end{tabular}
\end{center}
\caption{Flavor wave functions in Jaffe and Wilczek's model
\cite{jaffe}.
$[q_1q_2][q_3q_4]_+=\sqrt{\frac{1}{2}}([q_1q_2][q_3q_4]+[q_3q_4][q_1q_2])$
or $[q_1q_2]^2=[q_1q_2][q_1q_2]$ is the diquark-diquark part.
}\label{tab2}
\end{table*}

The magnetic moment of a compound system is the sum of the
magnetic moments of its constituents including spin and orbital
contributions,
\begin{equation}
\overrightarrow{\mu}=\sum\limits_{i}\overrightarrow{\mu_i}
=\sum\limits_{i}(g_{i}\overrightarrow{s_i}+\overrightarrow{l_i})
\mu_i,
\end{equation}
where $g_{i}$ is the $g$-factor of $i$-th constituent and $\mu_i$
is the magneton of the $i$-th constituent
\begin{equation}
\mu_i = \frac{e_i}{2m_i}\; .
\end{equation}

Treating the scalar diquarks as composite bosons, the pentaquark
can be regarded as a three-body system consisting of two scalar
bosons and a spin-$\frac12$ fermion. Thus
\begin{eqnarray}
\overrightarrow{\mu}&=& (g_1 \overrightarrow{0} +
\overrightarrow{l_1}) \mu_1 +(g_2 \overrightarrow{0} +
\overrightarrow{l_2}) \mu_2 +(g_3 \overrightarrow{\frac12} +
\overrightarrow{0}) \mu_3,
\nonumber\\
&=& \overrightarrow{l_1} \mu_1 +\overrightarrow{l_2} \mu_2 +g_3
\overrightarrow{\frac12} \mu_3 \; .
\end{eqnarray}
It is clear from the above equation that the pentaquark magnetic
moment arises from the spin of the antiquark and the relative
angular momentum between the two scalar diquarks which are
labelled as 1 and 2.

The orbital magnetic moment $\mu_l$ of a pentaquark in this model
is defined as
\begin{equation}
\mu_l \overrightarrow{l}= \overrightarrow{l_1} \mu_1 +
\overrightarrow{l_2} \mu_2\; .
\end{equation}
The spatial part of the diquark-diquark wave function can be
separated into the overall center-of-mass motion and the relative
motion
\begin{equation}
\Psi_{spatial}=\Psi({\bf R}) \psi({\bf r}),
\end{equation}
where
\begin{equation}
{\bf R} = \frac{m_1 {\bf r}_1 + m_2 {\bf r}_2}{m_1 + m_2}
\end{equation}
is the position of the center of mass and
\begin{equation}
{\bf r} = {\bf r}_1 - {\bf r}_2
\end{equation}
is the relative position between the two diquarks.  Now we
calculate each diquark's orbital angular momentum.

\begin{equation}
{\bf l}_1={\bf r}_1 \times {\bf p}_1 =({\bf
R}+\frac{m_2}{m_1+m_2}{\bf r}) \times (\frac{m_1}{m_1+m_2}{\bf
P}+{\bf p}).
\end{equation}

Since the pentaquark is a bound states of multi-quarks, the
average momentum of every quark is zero: $\langle {\bf p_i}
\rangle= 0$. So we have $\langle {\bf P} \rangle = 0$ and $\langle
{\bf p} \rangle =0$, from which we get:
\begin{eqnarray}
\langle {\bf l}_1 \rangle &=&\frac{m_2}{m_1+m_2} \langle {\bf
r}\times {\bf p} \rangle
+\frac{m_1}{m_1+m_2} \langle {\bf R}\times {\bf P} \rangle \nonumber \\
&=&\frac{m_2}{m_1+m_2} \langle {\bf l} \rangle
+\frac{m_1}{m_1+m_2} \langle {\bf L} \rangle .
\end{eqnarray}
${\bf l}={\bf r}\times {\bf p}$ is just the diquark-diquark's
relative orbital angular momentum.  ${\bf L}={\bf R}\times {\bf
P}$ is the diquark-diquark's orbital angular momentum for the
motion of center of mass.  The pentaquark states are assumed to
have only a relative angular momentum $l=1$ between the diquarks.
Thus $\langle {\bf L} \rangle=0$, which is manifest in the center
of mass frame. Therefore, we get
\begin{equation}
\langle {\bf l}_1 \rangle = \frac{m_2}{m_1+m_2} \langle {\bf l}
\rangle.
\end{equation}
\begin{equation}
\langle {\bf l}_2 \rangle = \frac{m_1}{m_1+m_2} \langle {\bf l}
\rangle.
\end{equation}
Now for the magnetic moment of the two diquark system, we have
\begin{eqnarray}\nonumber
\mu_l \overrightarrow{l} &=& \frac{m_2 \mu_1}{m_1+m_2}
\overrightarrow{l} +\frac{m_1 \mu_2}{m_1+m_2}
\overrightarrow{l}\;, \\
\mu_l &=& \frac{m_2 \mu_1}{m_1+m_2} +\frac{m_1 \mu_2}{m_1+m_2}.
\end{eqnarray}
Finally the magnetic moment of a pentaquark reads
\begin{eqnarray}
\mu&=& \langle 2\mu_{\bar{q}}\overrightarrow{\frac{1}{2}}
+\mu_l \overrightarrow{l} \rangle (J_z = \frac12) \nonumber\\
   &=& \left( \langle 1 0 \frac12 \frac12 \mid \frac12 \frac12 \rangle^2
- \langle 1 1 \frac12 -\frac12 \mid \frac12 \frac12 \rangle^2
\right) \mu_{\bar{q}} + \langle 1 1 \frac12 -\frac12 \mid \frac12
\frac12 \rangle^2 \mu_l
\end{eqnarray}

We present the expressions of the pentaquark magnetic moments in
Table \ref{tab3} and \ref{tab4}. We use $m_u=m_d=360$ MeV,
$m_s=500$ MeV and two sets of diquark mass parameters to compute
the magnetic moments of these $J^P=\frac{1}{2}^+$ pentaquarks. The
numerical results are summarized in Table \ref{tab5}.

\begin{table*}
\begin{center}
\begin{tabular}{ll}\hline
($Y,I,I_3$)          & Anti-decuplet Magnetic Moments \\
\hline
(2,0,0)&$\frac{1}{9}(\frac{1}{m_{ud}}-\frac{1}{2m_s})e_0$\\
(1,$\frac{1}{2}$,$\frac{1}{2}$)&$\frac{1}{27}[\frac{2}{m_{ud}+m_{us}}(\frac{m_{us}}{m_{ud}}+\frac{m_{ud}}{m_{us}})+\frac{1}{m_{ud}}-\frac{1}{2m_d}-\frac{1}{m_s}]e_0$\\
(1,$\frac{1}{2}$,-$\frac{1}{2}$)&$\frac{1}{27}[\frac{2}{m_{ud}+m_{ds}}(\frac{m_{ds}}{m_{ud}}-\frac{2m_{ud}}{m_{ds}})+\frac{1}{m_{ud}}+\frac{1}{m_u}-\frac{1}{m_s}]e_0$\\
(0,1,1)&$\frac{1}{27}[\frac{2}{m_{ud}+m_{us}}(\frac{m_{us}}{m_{ud}}+\frac{m_{ud}}{m_{us}})+\frac{1}{m_{us}}-\frac{1}{2m_s}-\frac{1}{m_d}]e_0$\\
(0,1,0)&$\frac{1}{54}[\frac{2}{m_{ud}+m_{us}}(\frac{m_{us}}{m_{ud}}+\frac{m_{ud}}{m_{us}})-\frac{2}{m_{ud}+m_{ds}}(\frac{2m_{ud}}{m_{ds}}-\frac{m_{ds}}{m_{ud}})$\\
      & \multicolumn{1}{r}{ $-\frac{2}{m_{ds}+m_{us}}(\frac{2m_{us}}{m_{ds}}-\frac{m_{ds}}{m_{us}})-\frac{1}{m_d}+\frac{2}{m_u}-\frac{1}{m_s}]e_0$}\\
 (0,1,-1)&$\frac{1}{27}[\frac{2}{m_{ud}+m_{ds}}(\frac{m_{ds}}{m_{ud}}-\frac{2m_{ud}}{m_{ds}})-\frac{2}{m_{ds}}+\frac{2}{m_u}-\frac{1}{2m_s}]e_0$\\
(-1,$\frac{3}{2}$,$\frac{3}{2}$)&$\frac{1}{9}(\frac{1}{m_{us}}-\frac{1}{2m_d})e_0$\\
(-1,$\frac{3}{2}$,$\frac{1}{2}$) &$\frac{1}{27}[\frac{2}{m_{us}+m_{ds}}(\frac{m_{ds}}{m_{us}}-\frac{2m_{us}}{m_{ds}})+\frac{1}{m_{us}}+\frac{1}{m_u}-\frac{1}{m_d}]e_0$\\
(-1,$\frac{3}{2}$,-$\frac{1}{2}$)&$\frac{1}{27}[\frac{2}{m_{us}+m_{ds}}(\frac{m_{ds}}{m_{us}}-\frac{2m_{us}}{m_{ds}})-\frac{2}{m_{ds}}+\frac{2}{m_u}-\frac{1}{2m_d}]e_0$\\
(-1,$\frac{3}{2}$,-$\frac{3}{2}$)&$\frac{1}{9}(\frac{1}{m_u}-\frac{2}{m_{ds}})e_0$\\
\hline
\end{tabular}
\end{center}
\caption{Expressions of the magnetic moments of the anti-decuplet
pentaquarks in Jaffe and Wilczek's model \cite{jaffe}, where $e_0$
is the charge unit.}\label{tab3}
\end{table*}

\begin{table*}
\begin{center}
\begin{tabular}{ll}\hline
($Y,I,I_3$)        &  Octet Pentaquark Magnetic Moments \\ \hline
(1,$\frac{1}{2}$,$\frac{1}{2}$)&$\frac{1}{27}[\frac{1}{m_{ud}+m_{us}}(\frac{m_{us}}{m_{ud}}+\frac{m_{ud}}{m_{us}})+\frac{2}{m_{ud}}-\frac{1}{m_d}-\frac{1}{2m_s}]e_0$\\
(1,$\frac{1}{2}$,-$\frac{1}{2}$)&$\frac{1}{27}[\frac{1}{m_{ud}+m_{ds}}(\frac{m_{ds}}{m_{ud}}-\frac{2m_{ud}}{m_{ds}})+\frac{2}{m_{ud}}+\frac{2}{m_u}-\frac{1}{2m_s}]e_0$\\
(0,1,1)&$\frac{1}{27}[\frac{1}{m_{ud}+m_{us}}(\frac{m_{us}}{m_{ud}}+\frac{m_{ud}}{m_{us}})+\frac{2}{m_{us}}-\frac{1}{m_s}-\frac{1}{2m_d}]e_0$\\
(0,1,0)&$\frac{1}{54}[\frac{1}{m_{ud}+m_{us}}(\frac{m_{us}}{m_{ud}}+\frac{m_{ud}}{m_{us}})+\frac{1}{m_{ud}+m_{ds}}(\frac{m_{ds}}{m_{ud}}-\frac{2m_{ud}}{m_{ds}})$\\
       &\multicolumn{1}{r}{$+\frac{4}{m_{ds}+m_{us}}(\frac{m_{ds}}{m_{us}}-\frac{2m_{us}}{m_{ds}})+\frac{1}{m_u}-\frac{1}{2m_d}-\frac{2}{m_s}]e_0$}\\
(0,1,-1)&$\frac{1}{27}[\frac{1}{m_{ud}+m_{ds}}(\frac{m_{ds}}{m_{ud}}-\frac{2m_{ud}}{m_{ds}})-\frac{4}{m_{ds}}+\frac{1}{m_u}-\frac{1}{m_s}]e_0$\\
(-1,$\frac{1}{2}$,$\frac{1}{2}$)&$\frac{1}{27}[\frac{1}{m_{us}+m_{ds}}(\frac{m_{ds}}{m_{us}}-\frac{2m_{us}}{m_{ds}})+\frac{2}{m_{us}}+\frac{2}{m_u}-\frac{1}{2m_d}]e_0$\\
(-1,$\frac{1}{2}$,-$\frac{1}{2}$)&$\frac{1}{27}[\frac{1}{m_{us}+m_{ds}}(\frac{m_{ds}}{m_{us}}-\frac{2m_{us}}{m_{ds}})-\frac{4}{m_{ds}}+\frac{1}{m_u}-\frac{1}{m_d}]e_0$\\
(0,0,0)&$\frac{1}{18}[\frac{1}{m_{ud}+m_{us}}(\frac{m_{us}}{m_{ud}}+\frac{m_{ud}}{m_{us}})+\frac{1}{m_{ud}+m_{ds}}(\frac{m_{ds}}{m_{ud}}-\frac{2m_{ud}}{m_{ds}})+\frac{1}{m_u}-\frac{1}{2m_d}]e_0$\\
\hline
\end{tabular}
\end{center}
\caption{Expressions of the magnetic moments of the octet
pentaquarks in Jaffe and Wilczek's model \cite{jaffe}, where $e_0$
is the charge unit.}\label{tab4}
\end{table*}

\begin{table*}[h]
\begin{center}
\begin{tabular}{c|c|c||c|c|c}\hline
 State    & \multicolumn{2}{|c||}{$\bf{\bar{10}}$}&State&\multicolumn{2}{|c}{\bf{8}}\\
 \hline
 ($Y,I,I_3$)                  & set I &set II &($Y,I,I_3$)        & set I &set II\\
 \hline
(2,0,0)                     &0.08 &0.29 &$-$&$-$&$-$  \\
(1,$\frac{1}{2}$,$\frac{1}{2}$) &0.037&0.22&(1,$\frac{1}{2}$,$\frac{1}{2}$)&0.018&0.21\\
(1,$\frac{1}{2}$,-$\frac{1}{2}$)&0.12 &0.22&(1,$\frac{1}{2}$,-$\frac{1}{2}$)&0.50 &0.65\\
(0,1,1)                     &-0.009&0.14&(0,1,1)&0.007&0.14\\
(0,1,0)                     &0.06 &0.12&(0,1,0)&-0.13&-0.13\\
(0,1,-1)                    &0.13 &0.09&(0,1,-1)&-0.27&-0.41\\
(-1,$\frac{3}{2}$,$\frac{3}{2}$)&-0.06&0.06 &$-$&$-$&$-$\\
(-1,$\frac{3}{2}$,$\frac{1}{2}$)&0     &0   &(-1,$\frac{1}{2}$,$\frac{1}{2}$) &0.41 &0.46\\
(-1,$\frac{3}{2}$,-$\frac{1}{2}$)&0.06&-0.06&(-1,$\frac{1}{2}$,-$\frac{1}{2}$)&-0.35&-0.52\\
(-1,$\frac{3}{2}$,-$\frac{3}{2}$)&0.12&-0.12&$-$&$-$&$-$\\
$-$&$-$&$-$&(0,0,0) &0.25 &0.37\\
\hline
\end{tabular}
\end{center}
\caption{Numerical results of the magnetic moments of ${\bf \bar{
10}}$ and {\bf 8} pentaquarks in unit of $\mu_N$ in Jaffe and
Wilczek's model \cite{jaffe}. For set I we use $m_{ud}=720$ MeV,
$m_{us}=m_{ds}=900$ MeV from Ref. \cite{lipkin} and for set II we
use $m_{ud}=420$ MeV, $m_{us}=m_{ds}=600$ MeV from Ref.
\cite{jaffe}. }\label{tab5}
\end{table*}

For $\Theta^+(1540)$, we have $\mu_{\Theta^+}=0.081{e_0\over
2m_N}\simeq 0.13\frac{e_0}{2M_{\Theta^+}}$ with Set I parameters
from Ref. \cite{lipkin}, while $\mu_{\Theta^+}\simeq
0.47\frac{e_0}{2M_{\Theta^+}}$ with Set II parameters from Ref.
\cite{jaffe}. $e_0$ is the positive charge unit. For the newly
observed pentaquark $\Xi_5^{--}(1860)$ which lies at
(-1,-$\frac{3}{2}$) of $I_3$-$Y$ diagram and is composed of
$ddss\bar{u}$,  we have $\mu_{\Xi_5^{--}}=0.116{e_0\over
2m_N}\simeq 0.23\frac{e_0}{2M_{\Xi_5^{--}}}$ with set I masses,
while $\mu_{\Xi_5^{--}}\simeq -0.23\frac{e_0}{2M_{\Xi_5^{--}}}$
with set II masses.

The magnetic moments are quite sensitive to the diquark mass
parameters. In our case, the pentaquark wave functions are
constructed using symmetries only. The physical particle may be
the mixture of those states at the same position in $I_3$-$Y$
diagram. For simplicity we have ignored all possible mixing
effects. Experimentally, it will be very challenging to identify
those pentaquarks without exotic quantum numbers.

\section{Pentaquark As A Bound State of One Scalar Diquark, One Tensor Diquark and One Anti-quark}\label{sec3}

Replacing the pair of scalar diquarks with one scalar and one
tensor diquark, Shuryak and Zahed argued that the pentaquark mass
may be lowered in this way \cite{shuryak}. The reason is that the
orbital excitation between diquarks is no longer necessary,
although the tensor diquark mass is slightly higher than the
scalar one. The orbital excitation between the up and down quarks
inside the tensor diquark ensures that the overall parity of the
$\Theta^+$ particle is positive. They claimed that they can get
much lighter states than in JW's model.

Now the tensor diquark is in the antisymmetric $\bf{\bar{3}_c}$
representation and its flavor is $\bf{\bar{3}_f}$. To get an $I=0$
$\Theta^+$, the diquark-diquark must be still in the symmetric
flavor $\bf{\bar{6}_f}$ representation. Thus the flavor wave
functions of the pentaquark states remain the same as in JW's
model after adding the single anti-quark.  However, the total
angular momentum of the tensor diquark is chosen to be $J=1$. The
Pauli principle demands that there is at least an orbital
excitation $l=1$ between two quarks inside the tensor diquark.
Thus the total spin of the two quarks inside the tensor diquark is
one. So the parity of the tensor quark is negative. As a result,
all of the eighteen pentaquark states have positive parity as in
JW's model while the spin of the pentaquark is either
$J=\frac{1}{2}$ or $\frac{3}{2}$.

The magnetic moment of a pentaquark in SZ's model is
\begin{eqnarray}
\overrightarrow{\mu}&=& (g_1 \overrightarrow{0} +
\overrightarrow{0}) \mu_1 +(g_2 \overrightarrow{1} +
\overrightarrow{0}) \mu_2 +(g_3 \overrightarrow{\frac12} +
\overrightarrow{0}) \mu_3,
\nonumber\\
&=& g_2 \overrightarrow{1} \mu_2 +g_3 \overrightarrow{\frac12}
\mu_3 \; .
\end{eqnarray}
The magnetic moment of a pentaquark comes from the spin of the
antiquark and the spin of the tensor diquark which is labelled 2.

The tensor diquark's $g$-factor can be calculated  from its two
components of an $i$- quark and a $j$- quark
\begin{equation}
g_2 \overrightarrow{1} \mu_2 = (g \overrightarrow{\frac12} +
\overrightarrow{l_i}) \mu_i +(g \overrightarrow{\frac12} +
\overrightarrow{l_j}) \mu_j,
\end{equation}
where $g=2$ is the quark $g$ factor.  The orbital magnetic moment
$\mu_l$ of the tensor diquark in this model is
\begin{equation}
\mu_l \overrightarrow{l}= \overrightarrow{l_i} \mu_i +
\overrightarrow{l_j} \mu_j,
\end{equation}
\begin{equation}
\mu_l = \frac{m_j \mu_i}{m_i+m_j} +\frac{m_i \mu_j}{m_i+m_j}.
\end{equation}
The $g$ factor of the tensor diquark can be evaluated with
\begin{equation}
g_2 \overrightarrow{1} \mu_2 = 2 \overrightarrow{\frac12} \mu_i +2
\overrightarrow{\frac12} \mu_j + \overrightarrow{1} \mu_l\; .
\end{equation}
\begin{equation}
g_2 \mu_2 = \langle 1 1 1 0 \mid 1 1 \rangle^2 (\mu_l +
\mu_i+\mu_j).
\end{equation}

We list the $g$ factor expressions of the tensor diquarks in SZ's
model below:
\begin{eqnarray}
g_{ud}&=&\frac{m_{ud}}{2}[\frac{2}{m_u}-\frac{1}{m_d}+\frac{1}{m_u+m_d}(\frac{2m_d}{m_u}-\frac{m_u}{m_d})],\\
g_{us}&=&\frac{m_{us}}{2}[\frac{2}{m_u}-\frac{1}{m_s}+\frac{1}{m_u+m_s}(\frac{2m_s}{m_u}-\frac{m_u}{m_s})],\\
g_{ds}&=&\frac{m_{ds}}{4}[\frac{1}{m_d}+\frac{1}{m_s}+\frac{1}{m_d+m_s}(\frac{m_s}{m_d}+\frac{m_d}{m_s})]
\end{eqnarray}
 $\Theta^+$'s magnetic moment in this case is:
\begin{eqnarray}
\mu_{\Theta^+} &=&\langle
2\mu_{\bar{q}}\overrightarrow{\frac{1}{2}}
+g_2\mu_2 \overrightarrow{1} \rangle \nonumber\\
   &=& \left( \langle 1 0 \frac12 \frac12 \mid \frac12 \frac12 \rangle^2
- \langle 1 1 \frac12 -\frac12 \mid \frac12 \frac12 \rangle^2
\right) \mu_{\bar{q}} + \langle 1 1 \frac12 -\frac12 \mid \frac12
\frac12 \rangle^2 g_2 \mu_2
\end{eqnarray}

\begin{table*}
\begin{center}
\begin{tabular}{ccc} \hline
 ($Y,I,I_3$)          & Decuplet Magnetic Moments & Numerical results\\ \hline
(2,0,0)&$\frac{1}{9}(g_{ud}\frac{1}{m_{ud}}-\frac{1}{2m_s})e_0$&0.23\\
(1,$\frac{1}{2}$,$\frac{1}{2}$)&$\frac{1}{27}(g_{ud}\frac{2}{m_{ud}}+g_{us}\frac{1}{m_{us}}-\frac{1}{m_s}-\frac{1}{2m_d})e_0$&0.26\\
(1,$\frac{1}{2}$,-$\frac{1}{2}$)&$\frac{1}{27}(g_{ud}\frac{2}{m_{ud}}-g_{ds}\frac{2}{m_{ds}}-\frac{1}{m_s}+\frac{1}{m_u})e_0$&0.09\\
(0,1,1)&$\frac{1}{27}(g_{us}\frac{2}{m_{us}}+g_{ud}\frac{1}{m_{ud}}-\frac{1}{m_d}-\frac{1}{2m_s})e_0$&0.30\\
(0,1,0)&$\frac{1}{27}(-g_{ds}\frac{2}{m_{ds}}+g_{us}\frac{1}{m_{us}}+g_{ud}\frac{1}{m_{ud}}+\frac{1}{m_u}-\frac{1}{2m_d}-\frac{1}{2m_s})e_0$&0.13\\
(0,1,-1)&$\frac{1}{27}(-g_{ds}\frac{4}{m_{ds}}+g_{ud}\frac{1}{m_{ud}}+\frac{2}{m_u}-\frac{1}{2m_s})e_0$&-0.04\\
(-1,$\frac{3}{2}$,$\frac{3}{2}$)&$\frac{1}{9}(g_{us}\frac{1}{m_{us}}-\frac{1}{2m_d})e_0$&0.33\\
(-1,$\frac{3}{2}$,$\frac{1}{2}$)&$\frac{1}{27}(g_{us}\frac{2}{m_{us}}-g_{ds}\frac{2}{m_{ds}}-\frac{1}{m_d}+\frac{1}{m_u})e_0$&0.16\\
(-1,$\frac{3}{2}$,-$\frac{1}{2}$)&$\frac{1}{27}(-g_{ds}\frac{4}{m_{ds}}+g_{us}\frac{1}{m_{us}}+\frac{2}{m_u}-\frac{1}{2m_d})e_0$&-0.01\\
(-1,$\frac{3}{2}$,-$\frac{3}{2}$)&$\frac{1}{9}(-g_{ds}\frac{2}{m_{ds}}+\frac{1}{m_u})e_0$&-0.17\\
\hline
\end{tabular}
\end{center}
\caption{Expressions of the magnetic moments of the anti-decuplet
pentaquarks and numerical results of them in unit of $\mu_N$ in
Shuryak and Zahed's model \cite{shuryak}.}\label{tab6}
\end{table*}

\begin{table*}
\begin{center}
\begin{tabular}{ccc}\hline
($Y,I,I_3$)        &  Octet Magnetic Moments& Numerical results \\
\hline
(1,$\frac{1}{2}$,$\frac{1}{2}$)&$\frac{1}{27}(g_{ud}\frac{5}{2m_{ud}}+g_{us}\frac{1}{2m_{us}}-\frac{1}{m_d}-\frac{1}{2m_s})e_0$&0.20\\
(1,$\frac{1}{2}$,-$\frac{1}{2}$)&$\frac{1}{27}(g_{ud}\frac{5}{2m_{ud}}-g_{ds}\frac{1}{m_{ds}}+\frac{2}{m_u}-\frac{1}{2m_s})e_0$&0.55\\
(0,1,1)&$\frac{1}{27}(g_{us}\frac{5}{2m_{us}}+g_{ud}\frac{1}{2m_{ud}}-\frac{1}{m_s}-\frac{1}{2m_d})e_0$&0.35\\
(0,1,0)&$\frac{1}{54}(g_{ud}\frac{1}{m_{ud}}-g_{ds}\frac{5}{m_{ds}}+g_{us}\frac{5}{2m_{us}}+\frac{1}{m_u}-\frac{2}{m_s}-\frac{1}{2m_d})e_0$&-0.07\\
(0,1,-1)&$\frac{1}{27}(g_{ud}\frac{1}{2m_{ud}}-g_{ds}\frac{5}{m_{ds}}+\frac{1}{m_u}-\frac{1}{m_s})e_0$&-0.50\\
(-1,$\frac{1}{2}$,$\frac{1}{2}$)&$\frac{1}{27}(g_{us}\frac{5}{2m_{us}}-g_{ds}\frac{1}{m_{ds}}+\frac{2}{m_u}-\frac{1}{2m_d})e_0$&0.68\\
(-1,$\frac{1}{2}$,-$\frac{1}{2}$)&$\frac{1}{27}(g_{us}\frac{1}{2m_{us}}-g_{ds}\frac{5}{m_{ds}}+\frac{1}{m_u}-\frac{1}{m_d})e_0$&-0.53\\
(0,0,0)&$\frac{1}{18}(g_{ud}\frac{1}{m_{ud}}-g_{ds}\frac{1}{m_{ds}}+g_{us}\frac{1}{2m_{us}}+\frac{1}{m_u}-\frac{1}{2m_d})e_0$&0.33\\
\hline
\end{tabular}\\
\end{center}
\caption{Expressions of the magnetic moments of the octet
pentaquarks and their numerical results in unit of $\mu_N$ in
Shuryak and Zahed's model \cite{shuryak}.}\label{tab7}
\end{table*}

We present the expressions and numerical results in Table
\ref{tab6} and \ref{tab7}. For quark masses, we use $m_u=m_d=360$
MeV, $m_s=500$ MeV. For the tensor diquark masses, we follow Ref.
\cite{shuryak} and use $m_{ud}^T=570$ MeV, $m_{us}^T=m_{ds}^T=720$
MeV. The scalar diquark mass is irrelevant in our calculation
because there is no orbital excitation among the three
constituents and the magnetic moment of the scalar diquark is
zero. The pentaquark $\Theta^+$'s and $\Xi_5^{--}$'s magnetic
moments are about $\mu_{\Theta^+}=0.37\frac{e_0}{2M_{\Theta^+}}$
and $\mu_{\Xi_5^{--}}=-0.35\frac{e_0}{2M_{\Xi_5^{--}}}$
respectively.

\section{Pentaquark As A Bound State of A Diquark and Triquark}\label{sec4}

An even simpler picture came from Karliner and Lipkin's diquark
triquark model \cite{lipkin}. In their model the two color
non-singlet clusters are kept apart by the P-wave angular momentum
barrier. Hence the color magnetic interaction occurs within two
clusters only. Between two clusters the color electric forces bind
them into a color singlet. The angular momentum barrier prevents
them from rearranging into the usual $KN$ system. The presence of
P-wave ensures the overall $\Theta^+$ parity is positive.

The two quarks in the triquark are in the symmetric ${\bf 6_c}$
representation. They couple with the antiquark to form an
$SU(3)_c$ triplet ${\bf 3}_c$.  The two quarks are in the
anti-symmetric flavor ${\bf \bar{3}_f}$ representation and the
triquark is in the symmetric ${\bf \bar 6}_f$. Thus the spin wave
function of the two quarks is symmetric. The spin of the triquark
is one half.

The direct product of the ${\bf \bar 3_f}$ of diquark and the
${\bf \bar 6_f}$ of triquark leads to ${\bf\bar{10}_f}$ and ${\bf
8_f}$ pentaquarks. There is one orbital angular momentum $L=1$
between the diquark and the triquark. The resulting $J^P$ of the
pentaquark is ${1\over 2}^+$. We list the flavor wave functions in
Table \ref{tab8} and \ref{tab9}.

\begin{table*}
\begin{center}
\begin{tabular}{ccc} \hline
($Y,I,I_3$)        &  Flavor wave functions     &Magnetic moments
\\ \hline
(2,0,0)                              &$[ud]\{ud\bar{s}\}$                                                          &   0.19   \\
(1,$\frac{1}{2}$,$\frac{1}{2}$)          &$\sqrt{\frac{1}{3}}([ud]\{ud\bar{d}\}+[ud]\{us\bar{s}\}+[us]\{ud\bar{s}\})$  &   0.17   \\
(1,$\frac{1}{2}$,-$\frac{1}{2}$)       &$\sqrt{\frac{1}{3}}([ud]\{ud\bar{u}\}+[ud]\{ds\bar{s}\}+[ds]\{ud\bar{s}\})$  &   -0.006  \\
(0,1,1)                   &$\sqrt{\frac{1}{3}}([us]\{ud\bar{d}\}+[us]\{us\bar{s}\}+[ud]\{us\bar{d}\})$             &   0.15   \\
(0,1,0)                   &$\sqrt{\frac{1}{6}}([us]\{ud\bar{u}\}+[us]\{ds\bar{s}\}+[ds]\{ud\bar{d}\}$              &   -0.03  \\
                            &$+[ds]\{us\bar{s}\}+[ud]\{us\bar{u}\}+[ud]\{ds\bar{d}\})$&\\
(0,1,-1)                &$\sqrt{\frac{1}{3}}([ds]\{ud\bar{u}\}+[ds]\{ds\bar{s}\}+[ud]\{ds\bar{u}\})$             &   -0.21  \\
(-1,$\frac{3}{2}$,$\frac{3}{2}$)       &$[us]\{us\bar{d}\}$                                                          &   0.13   \\
(-1,$\frac{3}{2}$,$\frac{1}{2}$)       &$\sqrt{\frac{1}{3}}([us]\{us\bar{u}\}+[us]\{ds\bar{d}\}+[ds]\{us\bar{d}\})$  &   -0.054  \\
(-1,$\frac{3}{2}$,-$\frac{1}{2}$)    &$\sqrt{\frac{1}{3}}([ds]\{ds\bar{d}\}+[ds]\{us\bar{u}\}+[us]\{ds\bar{u}\})$  &   -0.24  \\
(-1,$\frac{3}{2}$,-$\frac{3}{2}$)    &$[ds]\{ds\bar{u}\}$                                                            &   -0.43  \\
\hline
\end{tabular}
\caption{Wave functions and numerical results of the magnetic
moments of ${\bf \bar{10}}$ pentaquarks in unit of $\mu_N$ in
Karliner and Lipkin's model \cite{lipkin}. $Y$, $I$ and $I_3$ are
hypercharge, isospin and the third component of isospin
respectively. $\{q_1q_2\bar{q_3}\}\equiv[q_1q_2]\bar{q_3}$ is the
triquark's flavor wave function.}\label{tab8}
\end{center}
\end{table*}

\begin{table*}
\begin{center}
\begin{tabular}{ccc} \hline
($Y,I,I_3$)        &  Flavor wave functions     &Magnetic moments
\\ \hline
(1,$\frac{1}{2}$,$\frac{1}{2}$)      &$\sqrt{\frac{1}{6}}([ud]\{ud\bar{d}\}+[ud]\{us\bar{s}\})-\sqrt{\frac{2}{3}}[us]\{ud\bar{s}\}$  &   0.16    \\
(1,$\frac{1}{2}$,-$\frac{1}{2}$)     &$\sqrt{\frac{1}{6}}([ud]\{ud\bar{u}\}+[ud]\{ds\bar{s}\})-\sqrt{\frac{2}{3}}[ds]\{ud\bar{s}\}$  &   -0.14   \\
(0,1,1)                              &$\sqrt{\frac{1}{6}}([us]\{ud\bar{d}\}+[us]\{us\bar{s}\})-\sqrt{\frac{2}{3}}[ud]\{us\bar{d}\}$  &   0.16    \\
(0,1,0)                              &$\sqrt{\frac{1}{12}}([us]\{ud\bar{u}\}+[us]\{ds\bar{s}\}+[ds]\{ud\bar{d}\}$                    &   0.04    \\
                                       &$+[ds]\{us\bar{s}\})-\sqrt{\frac{1}{3}}([ud]\{us\bar{u}\}+[ud]\{ds\bar{d}\})$&\\
(0,1,-1)                             &$\sqrt{\frac{1}{6}}([ds]\{ud\bar{u}\}+[ds]\{ds\bar{s}\})-\sqrt{\frac{2}{3}}[ud]\{ds\bar{u}\}$  &   -0.07   \\
(-1,$\frac{1}{2}$,$\frac{1}{2}$)     &$\sqrt{\frac{1}{6}}([us]\{us\bar{u}\}+[us]\{ds\bar{d}\})-\sqrt{\frac{2}{3}}[ds]\{us\bar{d}\}$  &   -0.17   \\
(-1,$\frac{1}{2}$,-$\frac{1}{2}$)    &$\sqrt{\frac{1}{6}}([ds]\{ds\bar{d}\}+[ds]\{us\bar{u}\})-\sqrt{\frac{2}{3}}[us]\{ds\bar{u}\}$  &   -0.12   \\
(0,0,0)                              &$\frac{1}{2}([us]\{ud\bar{u}\}+[us]\{ds\bar{s}\}-[ds]\{ud\bar{d}\}-[ds]\{us\bar{s}\})$         &   -0.10   \\
\hline
\end{tabular}
\caption{Wave functions and numerical results of the magnetic
moments of ${\bf {8_f}}$ pentaquarks in unit of $\mu_N$ in
Karliner and Lipkin's model \cite{lipkin}. $Y$, $I$ and $I_3$ are
hypercharge, isospin and the third component of isospin
respectively.}\label{tab9}
\end{center}
\end{table*}

The intrinsic magnetic moment of the triquark is defined as
\begin{equation}
g_{tri} \mu_{tri} \overrightarrow{\frac12} = 2\mu_{q_1}
\overrightarrow{\frac{1}{2}}
+2\mu_{q_2}\overrightarrow{\frac{1}{2}}
+2\mu_{\bar{q}}\overrightarrow{\frac{1}{2}}.
\end{equation}
From the spin structure of the triquark, we get
\begin{equation}
\frac{1}{2}g_{tri}\mu_{tri}=\langle 1 1 \frac12 -\frac12 \mid
\frac12 \frac12 \rangle^2 (\mu_{q_1}+\mu_{q_2}) + \left(\langle 1
0 \frac12 \frac12 \mid \frac12 \frac12 \rangle^2 -\langle 1 1
\frac12 -\frac12 \mid \frac12 \frac12 \rangle^2 \right) \mu_{\bar
q}
\end{equation}

For the orbital part we have
\begin{equation}
\mu_l \overrightarrow{1} = \mu_{tri} \overrightarrow{l_{tri}}
+\mu_{di} \overrightarrow{l_{di}}
\end{equation}
\begin{equation}
\mu_l = \frac{m_{tri} \mu_{di} + m_{di} \mu_{tri}}{m_{tri} +
m_{di}},
\end{equation}
where $m_{di}$ is the mass of the diquark, $m_{tri}$ is the mass
of the triquark.

Now the magnetic moment of the pentaquark is
\begin{equation}
 \overrightarrow{\mu}
= \mu_l \overrightarrow{1} + g_{tri}\mu_{tri}
\overrightarrow{\frac{1}{2}}.
\end{equation}
\begin{equation}
\mu = \langle 1 1
\frac{1}{2}-\frac{1}{2}|\frac{1}{2}\frac{1}{2}\rangle^2 \mu_l
+\left( \langle 10
\frac{1}{2}\frac{1}{2}|\frac{1}{2}\frac{1}{2}\rangle^2- \langle 11
\frac{1}{2}-\frac{1}{2}|\frac{1}{2}\frac{1}{2} \rangle^2 \right)
\frac{1}{2}g_{tri}\mu_{tri}
\end{equation}

The results are listed in Table \ref{tab8} and \ref{tab9}. We use
$m_u=m_d=360$ MeV, $m_s=500$ MeV.  The mass of each diquark
(triquark) is the sum of its constituent mass as in Ref.
\cite{lipkin}.

\section{MIT bag model}\label{sec5}

Strottman discussed pentaquarks in the MIT bag model many years
ago \cite{strot}. All five quarks in a low-lying pentaquark are in
the same single particle orbital ground state with angular
momentum $l=0$. Thus they have negative parity. One can classify
the pentaquark states according to their spin, flavor and color
symmetry $SU(6) \otimes SU_f(3) \supset SU_s(2) \otimes SU(3)_c
\otimes SU(3)_f$ as \cite{strot}
\begin{equation}
|q^4 [f](\lambda \mu)_f (10)_c {\cal S}_q I_q Y_q \times \bar{q}
[1^5] (01)_f (01)_c \frac12 I_{\bar{q}} Y_{\bar{q}}; [f]_t (00)_c
SIY\rangle.
\end{equation}
The pentaquark state is constructed from a four quark cluster
$q^4$ and an anti-quark $\bar{q}$. Both the four quark cluster and
the single anti-quark are specified by their $SU(6)$
representation $[f]$, $SU_f(3)$ representation $(\lambda \mu)_f$,
$SU(3)_c$ representation $(\lambda \mu)_c$, and quantum numbers
spin $S_i$, isospin $I_i$ and hypercharge $Y_i$ respectively. The
$(00)_c$ color singlet pentaquark state is further specified by
its $SU(6)$ representation $[f]_t$, and quantum numbers spin
${\cal S}$, iso-spin $I$ and hypercharge $Y$.  Since the
anti-quark is in color $(01)_c$ representation, the four quark
cluster must be in $(10)_c$ to form a color singlet pentaquark
state. From the Pauli principle, the total wave function of the
four quark cluster must be anti-symmetric which requires that its
$SU_f(3)$ representation $(\lambda \mu)_f$ is conjugated to the
$SU(6)$ representation $[f]$. The calculation \cite{strot} showed
that all the ${\frac12}^-$ pentaquarks with mass less than 1900
MeV belong to the single $SU(6)$ representation $[21]$.

For the $\Theta^+$ particle, its quantum numbers are $(J,Y,I,I_3)
= (\frac12,2,0,0)$.  The lowest energy state was predicted around
1.70 GeV and has the four quark cluster in $SU(6)$ representation
$[22]$ (the conjugated $SU_f(3)$ representation of the four quark
cluster is $(02)_f$). With the Clebsch-Gordan coefficients we get
\cite{strot},
\begin{eqnarray} &&|q^4[22](02)_f I_q Y_q \times \bar{q} [1^5]
(01)_f I_{\bar{q}} Y_{\bar{q}};
[21] (00)_c \frac12 I Y \rangle \nonumber \\
&=&|q^4[22](02)_f (10)_c 1 I_q Y_q \times \bar{q} [1^5] (01)_f
(01)_c \frac12 I_{\bar{q}} Y_{\bar{q}}; (00)_c \frac12 I Y
\rangle.
\end{eqnarray}
Hence the spin of the four quark cluster must be $S=1$. As the
iso-spin of the four quark cluster in $\Theta^+$ is $T=0$, its
mangetic moment can be obtain from the average mangetic moment of
u- and d- quarks $\bar{\mu} = \frac12(\mu_u+\mu_d)$
\begin{equation}
\mu_{q^4} = 2 \bar{\mu} = \mu_u + \mu_d
\end{equation}
where $2$ is the $g$-factor of quarks. The pentaquark magnetic
moment is
\begin{eqnarray}
\mu_{\Theta^+} &=& \frac23 \mu_{q^4} - \frac13 \mu_{\bar{q}} \nonumber \\
&=& \frac{2e_0}{9m_u}-\frac{e_0}{9m_d}-\frac{e_0}{18m_s} =
\frac{e_0}{9m_u} - \frac{e_0}{18m_s}
\end{eqnarray}

For the $\Xi^{--}_5$ and $\Xi^{+}_5$ particle, their quantum
numbers are $(J,Y,I,I_3) = (\frac12, -1, \frac32, -\frac32)$ and
$(J,Y,I,I_3) = (\frac12, -1, \frac32, \frac32)$ respectively
\cite{strot}. The lowest energy state was predicted around 1.90
GeV with the four quark cluster still in $SU(6)$ representation
$[22]$. Similarly, we obtain
\begin{eqnarray}
\mu_{\Xi_5^{--}} &=& \frac23 (\mu_d+\mu_s) - \frac13 \mu_{\bar{u}} \nonumber \\
&=& -\frac{e_0}{9m_d}-\frac{e_0}{9m_s}+\frac{e_0}{9m_u} = -
\frac{e_0}{9m_s},
\end{eqnarray}
and
\begin{eqnarray}
\mu_{\Xi_5^{+}} &=& \frac23 (\mu_u+\mu_s) - \frac13 \mu_{\bar{d}} \nonumber \\
&=& \frac{2e_0}{9m_u}-\frac{e_0}{9m_s}-\frac{e_0}{18m_d} =
\frac{e_0}{6m_u} - \frac{e_0}{9m_s}.
\end{eqnarray}

If the four quark cluster in a pentaquark favors the $SU(6)$
representation $[22]$, the flavor of the four quark cluster will
be the conjugated $SU_f(3)$ representation $(02)_f$. The total
flavor symmetry of the pentaquark states will be
\begin{equation}
(02) \times (01) = (11) + (03),
\end{equation}
including an $\mathbf{8}_f$ and a $\bar{\mathbf{10}}_f$ (clearly,
the above exotic states belong to the $\bar{\mathbf{10}}_f$
representation). The magnetic moments of all these pentaquark
states are
\begin{equation}
\mu = \frac23 \mu_{q^4} - \frac13 \mu_{\bar{q}} ,
\end{equation}
with
\begin{equation}
\mu_{q^4} = 2 \bar{\mu} = \frac12 \sum \mu_{q_i}.
\end{equation}

\section{Discussions}\label{sec6}

Several theoretical papers have already discussed the $\Theta^+$
magnetic moments. In Ref. \cite{mag} Kim and Praszalowicz derived
relations for the anti-decuplet magnetic moments within the
framework of chiral soliton model in the chiral limit. The
$\Theta^+$ pentaquark magnetic moment is estimated to be $(0.2\sim
0.3) \mu_N$ \cite{mag}.

In Ref. \cite{hosaka},  Nam, Hosaka and Kim first estimated the
anomalous magnetic moment $\kappa_{\Theta^+}$ of the $\Theta^+$
pentaquark when they discussed the photoproduction of $\Theta^+$
from the neutron. For example, they have estimated
$\kappa_{\Theta^+} = - 0.7$ and $\kappa_{\Theta^+} = - 0.2$ for
$J^P={1\over 2}^+$ and $J^P={1\over 2}^-$ respectively using Jaffe
and Wilczek's diquark picture. Also, they have considered the
possibility of a K N bound system, with $\kappa(KN, J^P={1\over
2}^+) = - 0.4$ and $\kappa(KN, J^P={1\over 2}^-) =- 0.5$. These
numbers are all in unit of $\Theta^+$ magneton ${e_0\over
2m_{\Theta^+}}$.

A quark model calculation of the $\Theta^+$ magnetic moment using
Jaffe and Wilczek's model was also performed in Ref. \cite{zhao},
where Zhao arrived at
\begin{equation}
\mu_{\Theta^+}=0.13{e_0\over 2m_{\Theta^+}}
\end{equation}
for the positive parity $\Theta^+$ pentaquark. He also discussed
the possibility of $\Theta^+$ parity being negative. In this case
he estimated its magnetic moment as the sum of $(u \bar s)$ and
$(udd)$ clusters. Numerically
\begin{equation}
\mu_{\Theta^+}={e_0\over 6m_s} \; .
\end{equation}

We have estimated the magnetic moment of the $\Theta^+$ pentaquark
state in the framework of light cone QCD sum rules using the
photon distribution amplitude. We have obtained a quite small
value: $|\mu_{\Theta^+}|=(0.12\pm 0.06) \mu_N$ \cite{huang}.

\begin{table*}
\begin{center}
\begin{tabular}{c|c|c|c|c}
\hline
\multicolumn{2}{c}{ }\vline& $\Theta^+$ & $\Xi^{--}_5$&$\Xi^{+}_5$\\
\hline
\multicolumn{2}{c}{Ref. \cite{mag}}\vline&$0.2\sim 0.3$  &$-0.4$ &$0.2$ \\
\hline
\multicolumn{2}{c}{Ref. \cite{hosaka}}\vline&$0.2\sim 0.5$ &$-$ &$-$ \\
\hline
\multicolumn{2}{c}{Ref. \cite{zhao}}\vline &$0.08\sim 0.6$ &$-$ &$-$ \\
\hline
\multicolumn{2}{c}{Ref. \cite{huang}}\vline&$(0.12\pm 0.06)$ &$-$ &$-$ \\
\hline
        &Jaffe and Wilczek's model                  &0.08 &0.12 &-0.06 \\
\cline{2-5}
 Present&Shuryak and Zahed's model        &0.23 &-0.17&0.33 \\
\cline{2-5}
  Work  &Karliner and Lipkin's model    &0.19 &-0.43&0.13 \\
\cline{2-5}
         &Strottman's model       &0.37 &-0.42&0.45 \\
\hline

\end{tabular}
\end{center}
\caption{Comparison of magnetic moments of $\Theta^+$,
$\Xi^{--}_5$ and $\Xi^{+}_5$ from other group's calculation with
the present work using different pentaquark models in literature.
The numbers are in unit $\mu_N$.}\label{tab10}
\end{table*}

In Table \ref{tab10} we have collected all available predictions
of the magnetic moments for $\Theta^+$, $\Xi^{--}_5$ and
$\Xi^{+}_5$ pentaquarks. There may exist strong mixing between the
other $\bar{\mathbf{10}}_f$ members and the corresponding octet
pentaquarks. Experimentally it will be very difficult to
distinguish them. So we focus on the $\Theta^+$, $\Xi^{--}_5$ and
$\Xi^{+}_5$ pentaquarks which lie on the corners of the
$\bar{\mathbf{10}}_f$ triangle and admit no mixing. It's
interesting to note that different models tend to yield relatively
close numerical results for the $\Theta^+$ magnetic moment. But
their predictions for $\Xi^{--}_5$ and $\Xi^{+}_5$ vary
significantly, which in turn will affect the photo-production or
electro-production total and differential cross sections. With the
accumulation of events, pentaquark magnetic moments may be
extracted from the comparison of theoretical and experimental
cross sections eventually in the near future, which may help us
distinguish different models and deepen our understanding of the
underlying dynamics governing their formation.

This project was supported by the National Natural Science
Foundation of China under Grant 10375003, Ministry of Education of
China, FANEDD and SRF for ROCS, SEM.


\end{document}